\documentclass[review, 12pt]{elsarticle}

\usepackage{hyperref}
\usepackage{tikz}
\usepackage{scalefnt}
\usepackage{amsmath}

\newcommand{\set}[1]{\left\{#1\right\}}

\journal{Journal of \LaTeX\ Templates}

\usepackage{setspace}

\begin{document}

\begin{frontmatter}

\title{Community-Based Security for the Internet of Things}

\author{Quanyan Zhu}
\address{Department of Electrical and Computer Engineering, New York University, \\
2 Metrotech Center, Brooklyn, 11201, USA}

\author{Stefan Rass, Peter Schartner}
\address{Institute of Applied Informatics, System Security Group, Universitaet Klagenfurt, Universitaetsstrasse 65-67, 9020 Klagenfurt, Austria}

%
%
%
%

\begin{abstract}
With more and more devices becoming connectable to the internet, the number
of services but also a lot of threats increases dramatically. Security is
often a secondary matter behind functionality and comfort, but the problem
has already been recognized. Still, with many IoT devices being deployed
already, security will come step-by-step and through updates, patches and
new versions of apps and IoT software. While these updates can be safely
retrieved from app stores, the problems kick in via jailbroken devices and
with the variety of untrusted sources arising on the internet. Since
hacking is typically a community effort these days, security could be a
community goal too. The challenges are manifold, and one reason for weak or
absent security on IoT devices is their weak computational power. In this
chapter, we discuss a community based security mechanism in which devices
mutually aid each other in secure software management. We discuss
game-theoretic methods of community formation and light-weight
cryptographic means to accomplish authentic software deployment inside the
IoT device community.
\end{abstract}


\end{frontmatter}


\section{Introduction}
The potential of services offered by the future internet of things is
accompanied by an equally strong growing potential of new threats. The
internet of things induces the trend to turn special purpose devices like TVs,
radios, etc. into universal platforms able to execute arbitrary pieces of
software and with strong communication abilities. This new power makes them
vulnerable to the same types of malware that was previously only seen in
classical computer networks. One of them is ransomware, which although dating
back to the 1980s, sees a revival and is a major pillar of the cybercrime
ecosystem today \cite{Maor.2015}. Ransomware is typically a weapon against
the masses, and with the goal of pressing money. Advanced persistent threats
are on the opposite being highly targeted attacks against specific victims,
and typically not about monetary gain, but to demonstrate the power and to cause
maximal damage. The common denominator of both extremes, and most that live
between these two is their focus on the weakest element, which is typically
the human. Nature itself teaches that flocks (in general communities) have
much higher chances to survive than any of their individuals would have on
their own. Why not adopt and systematize this well-proven behavior in
security and the internet of things? Awareness of humans is a notoriously
volatile state since press and media have an undoubted power in sensitizing
people for a topic, but this regards every topic on which news are reported.
So, more recent news tend to supersede older ones and hence awareness about
threats are continuously fading away due to the mass of information that
people are confronted with every day.

In that sense, the internet of things and, more generally, the information
society itself creates one of the biggest dangers, by constantly overloading
individuals with information so that recognizing the real danger when it
occurs is harder than ever. Can we find a way out of this? There is surely no
easy answer, and societal changes towards better awareness should be expected
to solve the problem in near future. But human communities are not the only
ones that we can create, and the internet of things, which basically is a
community, offers much more controllable dynamics than any human society.

\begin{figure}
  \centering
  \scalefont{.7}
  \begin{tikzpicture}[scale=0.7]
    \node at (0,0) {\includegraphics[scale=0.7]{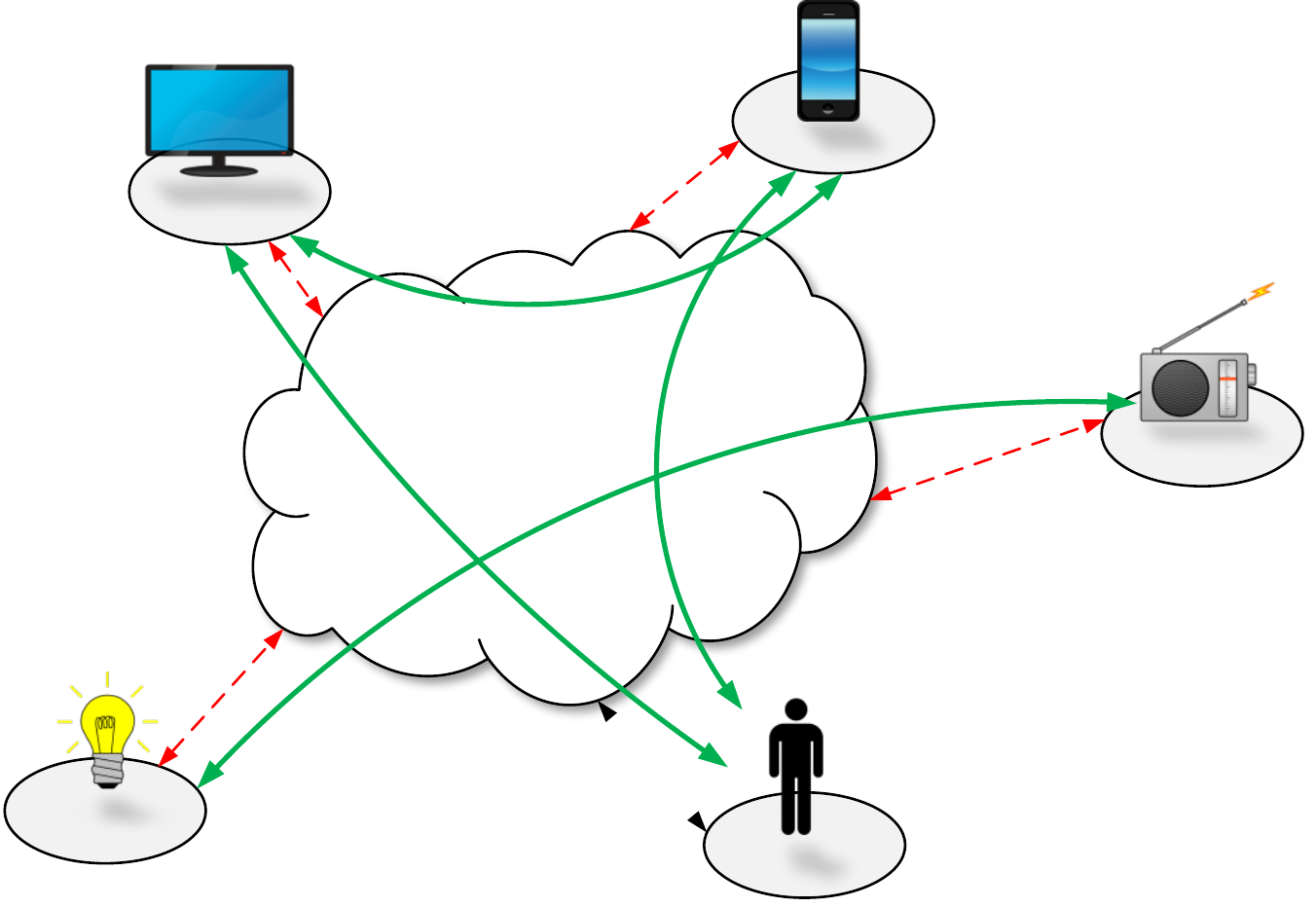}};
    \node at (-2,1) {internet of things};
    \node at (-1.5,0.5) {\shortstack[l]{$\to$ \textbf{smart}\\\textbf{community}}};
    \node[rotate=-47] at (-2.8,-0.5) {\textcolor[rgb]{0,0.5,0}{secure app sharing}};
    \node[rotate=21] at (3.5,-0.5) {\textcolor[rgb]{1,0,0}{\shortstack{(shared) threat\\warnings}}};
  \end{tikzpicture}
  \caption{Turning the IoT into a Smart Community}\label{fig:smart-community}
\end{figure}
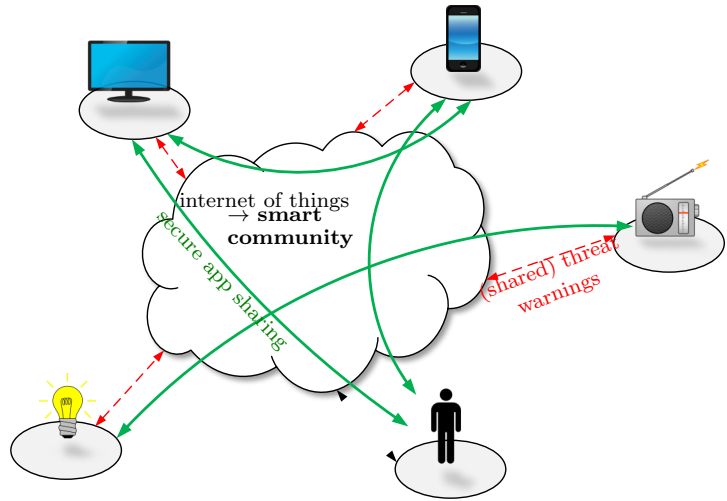

Speaking more concretely, let us look at a documented case of ransomware,
having locked a TV screen \cite{Cimpanu.2016}. The issue was apparently due
to the installation of a mysterious app. More critically, the viruses are
usually able to jump between different devices, partly also due to the strong
prevalence of only a few platforms. For example, Android is running on
tablets, mobile phones, eBook readers, TVs and many more. So, there is no
technical barrier for a virulent app to pass from one device to the other.
The links are exactly what IoT provides, so the world is open to any number
of viruses to be fruitful and multiply and infect the world. Indeed, why not
turn IoT devices themselves into a community in which individuals (devices)
mutually inform each other about threats, countermeasures and securely share
apps? Since trust of humans in technology seems to be high already, why not
let the awareness be up to the devices rather than the people? Figure
\ref{fig:smart-community} sketches the vision on this smart community living
in the IoT.

With apps obtainable from various sources, only one of which is the trusted
official app store, drive-by downloads have become a common way of injecting
malware on mobile platforms, and perhaps also future IoT. In a drive-by
download attack, the adversary obtains a legitimate copy of an app
installation file (on Android, this would be an apk-file), unpacks it (as it
is basically a compressed archive), adds the malware, and repacks everything
back into the harmlessly looking app installation file. The unaware victim
obtains the apk-file from some source and installs the app, but also the
malware that unknowingly ships with it. Official key stores run by platform
vendors (like Google, Apple, etc.) perform lots of screenings and security
checks, and a drive-by download is most unlikely for these sources. Still,
one attack vector on IoT and mobile platforms is to disconnect the user from
the official app store (say, by redirecting or blocking the connection
somehow), to enforce the user to look for alternative sources. There are
usually many of them, but not all being equally trustworthy.

More specifically, mobile platforms can be used with mobile device management
systems (MDMs) that let anyone (primarily enterprises) run their own app
stores, why not turn the community itself into a trusted app store for
itself? While app stores enjoy the luxury of cryptographic security (digital
signatures to verify the originality of the app), those cryptographic
security precautions are not easily established inside a loose community that
cannot perform a proper and decent key management.

Fortunately, there may be not even a strong need for a full-fledged public
key infrastructure to endow an IoT user community with the cryptographic
assurance of mutual authentication and hence trustworthiness in the shared
apps! Indeed, past research has come up with proposals on how to use keys
shared only with friends (or more general, the local neighborhood), to
establish end-to-end authentication (and security) even with strangers with
which no public key data is shared. The community, in turn, should be formed
so that everyone in it has an incentive to contribute, i.e., actively
communicate potential threat discoveries, and actively serve as a trusted
source of data, software, and information for others. The challenge herein is
building the community in the proper way to achieve (cryptographic) security
for everyone. This is two problems, one of community formation, the
other on security establishment. The community can be formed based on
incentives so that it is the best option for everyone in the community to
contribute to it. Suitable mechanisms to this end can be constructed from
game theory, as we will discuss later in this chapter.

\section{State of The Art on App Security}
Taking mobile platforms as templates as to how IoT devices may look like in
future, the security of mobile devices rests on four main pillars:

\begin{enumerate}
  \item Screening of apps at app stores: before an app goes online for
      download, most app stores scan for malicious code or patterns
      (similar to what a virus scanner would do).
  \item Access restrictions and permissions by the operating systems: when
      an app is installed, usually the required access privileges are
      displayed so that the user can explicitly consent. An informed
      decision hereby judges the combination of permissions and apps rather
      than each one individually, since the sum of permissions is typically
      much more powerful (dangerous) than each permission is on its own. At
      this point, care is up to the user (for example, in becoming
      suspicious if a torchlight app asks for network access; even if this
      app has no network access rights, it could share sensitive
      information via the clipboard with a ``befriended'' app from the same
      vendor that can send away the data). Newer versions of Android, for
      example, also allow to revoke such rights even after installation.
  \item App sandboxing: within the mobile device prevents apps from
      mutually accessing each other's memory space or the process itself.
      Still, there are several possibilities to let apps talk to each other
      and exchange data, which can be exploited maliciously (similar to how
      the heartbleed exploit worked for SSL).
  \item User awareness: the checks done at the app stores and everything
      that the operating system does hinges on the user acting carefully.
      This care includes an original (especially not jailbroken) copy of
      the platform, as well as a careful decision on which app to install.
      If the app's purpose is inconsistent with the privileges that it
      requires, then this is an indication of potential danger.
\end{enumerate}

The last pillar is where the community can help the most: it is not always
obvious why an app should not need some privilege, and some paid apps may
have free-of-charge siblings that are simply unofficial and may thus not be
found in an official app store. In both cases, community knowledge can aid
and guide the user. For example, if an app is reported (to the community) as
asking for strange privileges, then a warning could be issued to a new user.
Likewise, untrusted sources can be replaced by the community acting as a
team, thus making it more difficult for the adversary to spread its malicious
content.

Using opinions is already standard in app stores, where apps get user ratings
to help others find the best app for their needs. Establishing a similar
system for security appears as a natural next step, and can be achieved using
simple means. In fact, the cryptographic assurance that users get when
retrieving an app over official channels can – in a similar way – be provided
by (and to) the whole user community, by proper incentive and authentication
mechanisms; the former resting on game theory, the latter rooting in
cryptography. Hereafter, we will thus look at ways to safely distribute apps
with cryptographic assurance.

\section{Community Based Security}
The basic goal to accomplish for an attacker is tricking an unaware user into
installing an app from a distrusted source. While app stores can easily
certify the genuineness of their items by public key cryptography (digital
signatures), not all software is obtained from these stores, and certificate
management is typically a challenge of its own (despite rich and useful
theory behind it).

Normally, communities rely on one or more authorities for the authentic
distribution of software and threat information (warning) communication. Why
not have the community do that job collaboratively? The conventional approach
to authentication via public key cryptography would require a fully pervasive
public key management in all devices. While this is certainly possible, the
cryptographic operations are expensive (maybe too expensive for some IoT
devices) and the variety of vendors will probably render the entire
architecture quite complex. Symmetric cryptography, on the other hand, comes
light-weight and can be used for authentication based on keys that need to be
established only in the local neighborhood of a device. This simplifies
matters and takes out at least one of the central authorities towards a more
decentralized approach.

Also, the sharing of threat notifications and patches can be made up to the
community, in addition to central such news distribution mechanisms.
Essentially, the community can engage in multi-peer credibility checking
towards an early warning system about drive-by downloads.

\subsection{Multi-Peer Credibility Checking}

A drive-by download occurs if an installation package is modified by an
attacker to contain malicious code additionally to the actual (legitimate) app code.
Adding the malware is the simple part, the tricky bit is getting someone else
to download the now malicious app. One way of enforcing this is disconnecting
people from the app store (temporarily), to enforce the search for alternate
sources for an app.

Assume that an app has been originally distributed from a trusted source in
first place, which is -- in any case -- the app store (hence, its role is
equally crucial also in a community based security approach, as it acts as
the initial point of trust before the decentralized security can come to
play). Once the app -- in its legitimate form -- has reached some outspread,
suppose a user wants to install it from a given installation package. The
open source domain offers a simple and effective security precaution in the
form of fingerprints: usually, an installation package is accompanied by
several mirrors and a (cryptographic) hash-sum (often, MD5 or SHA-1). The
genuineness of the downloaded file can be checked by verifying the checksum
on the website. For this to work, the checksum and installation package
should come from different sources, so that the confidence in the
verification can be based on the unlikely event of the attacker having
conquered both, the data and the checksum host(s) at the same time.

Some software systems (like the statistical software R or the package manager
Chocolatey) by default do such a verification, which is no more expensive
than one hash-and-compare operation. IoT devices will not have much
difficulties in offering the necessary computational power to achieve
security by the same technique; relying on the community.

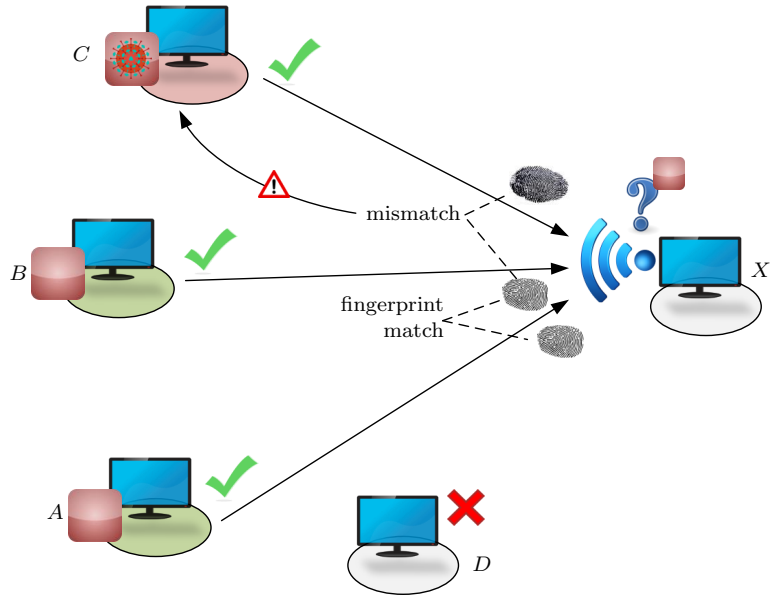
\begin{figure}
  \centering
  \scalefont{0.7}
  \begin{tikzpicture}[scale=0.7]
    \node at (0,0) {\includegraphics[scale=0.7]{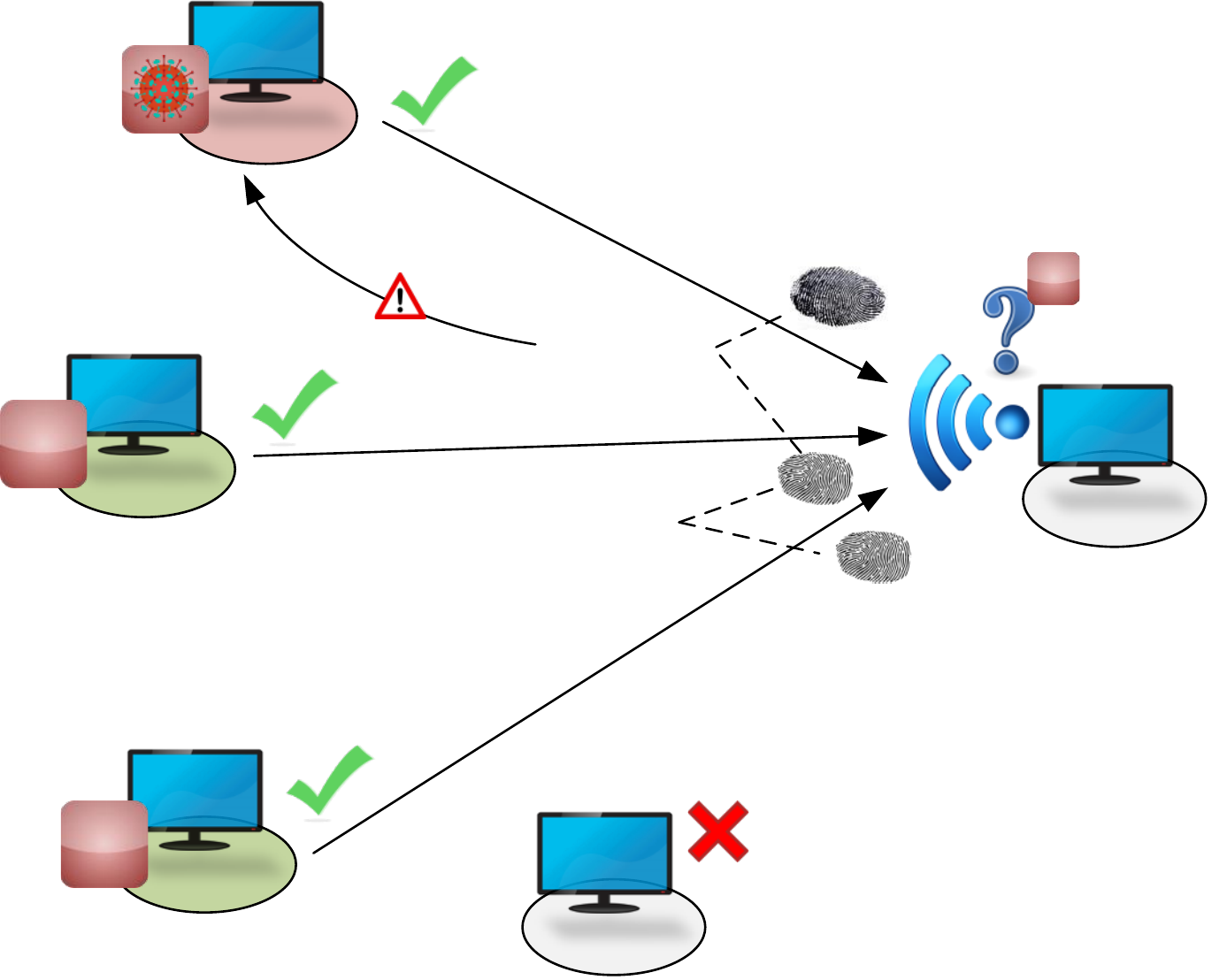}};
    \node at (0.2,1.68) {mismatch};
    \node at (-0.2,-0.3) {\shortstack[r]{fingerprint\\match}};
    \node at (-6.1,4.7) {$C$};
    \node at (-7.3,0.54) {$B$};
    \node at (-6.6, -4) {$A$};
    \node at (1.5, -5) {$D$};
    \node at (6.8, 0.6) {$X$};
  \end{tikzpicture}
  \caption{Multipeer Credibility Checking}\label{fig:multipeer-credibility-checking}
\end{figure}

Figure \ref{fig:multipeer-credibility-checking} gives an example of this
process, where a user $X$ seeks to install some app on a smart TV. To get the
installation package, it issues a call-out asking for others having installed
the app on their (similar) devices. Suppose devices $A,B$ and $C$ have the
app, while device $D$ sees the request but must remain silent since the app
is not available on this device. The peers $A,B$ and $C$, however, respond by
sending back a fingerprint (hash checksum) of their locally installed apps.
Among the three incoming fingerprints, those from $B$ and $C$ match, while
that of $A$ is different. This is already an indication that either $B$ and
$C$, or $A$, must have some malicious version of the app installed. Going
with a majority vote (based on the hypothesis that less infections are more
likely than a pandemic outbreak of the malware), the user would thus request
the app from either $B$ or $C$ (making a random choice), and inform $A$ about
its suspicion to have malware there. In this way, $A$ gets informed, and can
– upon next possibility – retrieve either a clean version of the app from the
store, or communicate this fingerprint to others to check their own copies of
the app (similar to a signature-based virus scan). Similar techniques are
also deployed for public key management, such as for trust management in PGP
and GPG \cite{Callas.2007,Penning.2017}.

Once the user has found a source to get the app from, it must authentically
retrieve it. This can be done by multipath authentication
\cite{Rass&Schartner2010}.

\subsection{Incentive Based Community Formation} It is apparent that the
credibility checking proposed relies on community building, since IoT peers
need to establish a trusted neighborhood to ask for credibility and also to
help with the authentication if software shall be deployed. Mechanisms to
form these trust clusters can be understood using agent-based models and
game-theoretic tools. The formation of the community is dynamic and arises
from a network formation process or a cooperative game. The formation of a
trustworthy IoT community plays an essential role in enabling the
collaboration among devices. The community formation is naturally a dynamic
problem in which nodes can join and depart the network and form communities
of their interest. This process can be modeled using a network formation
process in which each node is characterized by its type, preferences, trust,
and utility. The type of the nodes refers to the functionalities, the age and
the version of the devices. The devices of the same type often benefit more
from the collaboration. However, devices of different types can also learn
about emerging cyber threats and protection mechanisms from devices of
similar types. Each device has a preference over the devices that it wants to
communicate with. The preferences are dependent on the trustworthiness of the
information received from other nodes and the utility of the collaboration.
Based on the preferences, each node can choose a subset of existing nodes to
initiate a request of connections and a link is formed between two nodes if
the request is accepted. The acceptance of a request can be determined by a
node through a cost-and-benefit evaluation; i.e., a node can determine the
collaboration devices through his own preferences. This request-and-approval
process is implemented at each node at every stage. As a result, a
large-scale network is formed in a distributed fashion. As the parameters of
the system change, nodes can terminate their collaboration with other nodes
and establish new collaborations over time. In addition, new nodes will join
the network and the existing nodes can leave the system. Hence the networks
formed are highly dynamic. The emergence of the community of the dynamic
network indicates the formation of the collaboration community. One important
phenomenon that we observe is the homophily, in which nodes of the same types
often form a community and share information together. Game-theoretic methods
can be used to form an agent-based model to predict the structure of the
network and the emergence of the communities by analyzing the Nash
equilibrium of the game.

Incentive mechanism design is an important tool to create incentives in the
network so that nodes actively share information with other nodes trustfully
and untrustworthy devices or free-riders will be isolated and disconnected
from the network. In the recent work of \cite{Zhu.2012}, mechanism design
tools enable the system to reach a desirable and unique Nash equilibrium that
allow devices to communicate in a conducive environment in which nodes
endeavor to contribute knowledge and resource to assist connected nodes in
the community. Any selfish or free-riding behavior will receive a tit-for-tat
response from the nodes in the community as a consequence. In this way,
healthy and growing collaborative communities will be achieved and
maintained. Mechanism design tools can also shape the size and the structure
of the network. With an appropriate design of incentive parameters, the
network can grow to a desirable size by encouraging the participations from
new devices. The connectivity and clustering of the community network can
also be controlled by creating supernodes that behave as information hubs and
incentivizing nodes to reach unconnected devices that can benefit from
joining the community.

\subsection{Multipath Authentication}

Let an IoT device have made contact with its neighborhood in the network,
i.e., any ``surrounding'' device that responds upon some sort of
``Hello-Message'' sent out upon the first connection (a direct connection
could, for example, use a time-to-live set to 1 in order to get only the
adjacent network devices; but farther distances are equally possible and
legitimate here).

Suppose that a new IoT device makes a handshake with its network neighbors,
and establishes a shared secret for later authentication. The pairing can be
done by any means, such as via firmware, Bluetooth, near field communication,
etc. Preferably, it can be even done at manufacturing time for a whole
production lot, so as to achieve a distribution in (geographic) proximity but
in highly diverse networks (different homes, distinct enterprises, etc.). In
(smart) homes, for example, this avoids all paired devices to communicate
over the same potentially compromised hub, since the paired devices are
located in different areas and under control of independent users. The mutual
finding of two IoT devices can, relative to network segmentation, firewalls
or other logical barriers, happen as a particular service on the application
layer, which needs to be allowed to run over the IoT. However, communicating
devices are the main purpose of IoT anyway.

In any case, note that unlike for general public key schemes, our schemes do
not require as frequent key updates over the lifetime as IoT device. For
mobile phones, as an example, an expected lifetime of 2 to 3 years (until it is
replaced by a more modern version) could make a key update even unnecessary
at all.

The idea of multipath authentication is the following: if $B$ wants to send
an authentic message to $A$, but has neither a shared key nor public-key
certificate from $A$, it employs its network neighbors to certify $B$'s
identity to $A$. To this end, $B$ shares a (distinct) key with each of its
neighbors, and attaches a set of message authentication codes (MACs) to the
data for $A$, indicating who $A$ should contact to have each MAC verified.
Upon reception of the data and MACs, $A$ can contact each neighbor of $B$ and
ask for a verification of the MACs. This validation process somewhat
resembles how handwritten signatures are verified in real life upon leaving a
signature sample that can be compared to the handwritten signature in
question. Electronically, the process can be run just alike, at the appeal of
coming cheap, since the most expensive operation is the key exchange (done
only once), but all subsequent operations being fast and efficient algorithms
from symmetric cryptography.

The security of the scheme, unlike that of public key cryptography, can be
made independent of unproven mathematical conjectures and rests only on the
assumption of a ``sufficiently small'' portion of the network having been
compromised. In that sense, the achievable security is ``unconditional''.
This avoids complicated assumptions that can make public key cryptography
somewhat opaque to people outside the expert community (thus, adding
negatively to the trust in these feelingly black boxes); see
\cite{Koblitz.2007} for an excellent introduction and discussion of the
issue. Finally, we remark that the way of proving security uses game-theory
at the core \cite{Rass&Schartner2010}.

\subsection*{Experimental Implementations}

\begin{figure}
  \centering
  \includegraphics[scale=0.7]{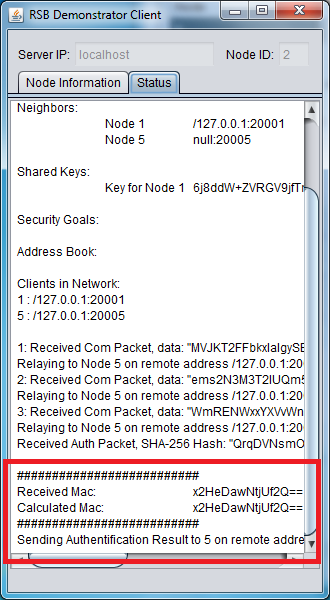}
  \caption{Multipath Authentication Prototype from \cite{Rass&Rainer2015}}\label{fig:mpa-prototype}
\end{figure}

The concept of multipath authentication has previously been implemented and
demonstrated to work on layer 7. In \cite{Rass&Rainer2015}, a demonstrator
has been reported that implements arbitrary secure end-to-end communication
(confidential and authentic) between devices where only locally paired ones
share a common secret (in an IoT setting, exchanging these is possible in
various ways, such as QR codes, near-field communication, etc.). The main
assumption upon which the security rests is multi-path source routing, where
paths do not intersect. Given a sufficiently dense network and accurate
information on the topology, it is not difficult to let the devices do the
routing on the application layer (as was shown in \cite{Rass&Rainer2015},
where a local Java client was used to handle these matters). Figure
\ref{fig:mpa-prototype} shows a screenshot of the past prototype, which is
(cryptographically) lightweight and implemented in Java to run on all
platforms. The screenshot shows parts of the log of a protocol run where node
2 was asked by node 5 to verify the MACs that node 2 received from node 1.
Node 5 shares a secret with node 1 so that it can confirm (see the ``boxed''
part on the bottom of the window) that the MAC it computed using the secret
shared with 1 matches what it received from 2 upon the authentication
request. Thus, node 1's ``signature'' is verified by node 2 and this is told
to node 5.

\section{Proposed Architecture}

\begin{figure}
  \centering
  \scalefont{0.7}
  \begin{tikzpicture}[scale=0.7]
    \node at (0,0) {\includegraphics[scale=0.7]{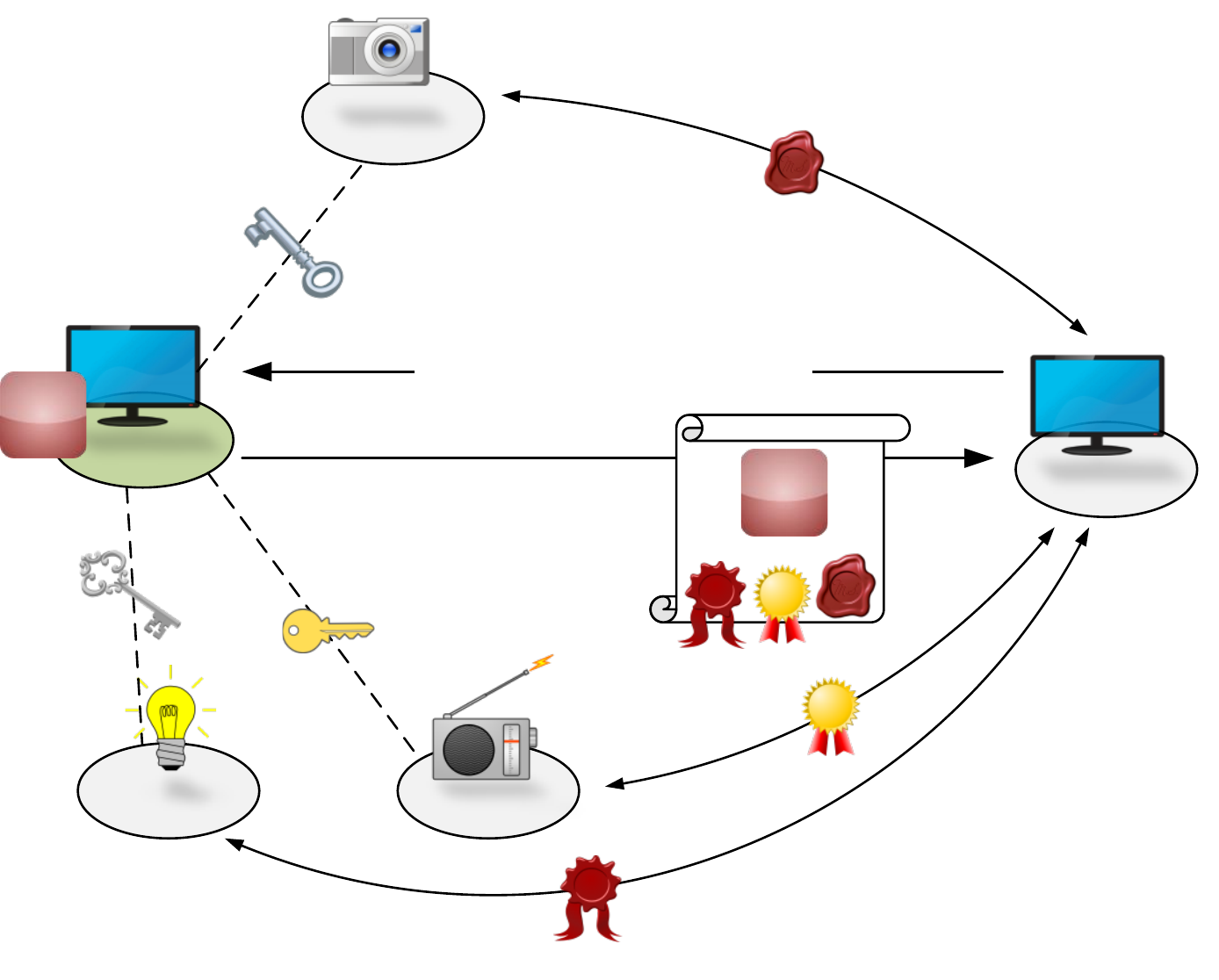}};
     \node at (-5.8,3.5) {\shortstack{shared keys\\(``neighborhood'')}};
    \node at (3,4.5) {\shortstack{MAC verification\\request}};
    \node at (0,1.4) {app download request};
    \node at (-0.9,-0.1) {\shortstack[r]{app file with\\3 MACs attached}};
  \end{tikzpicture}
  \caption{Multipath Authentication}\label{fig:mpa}
\end{figure}
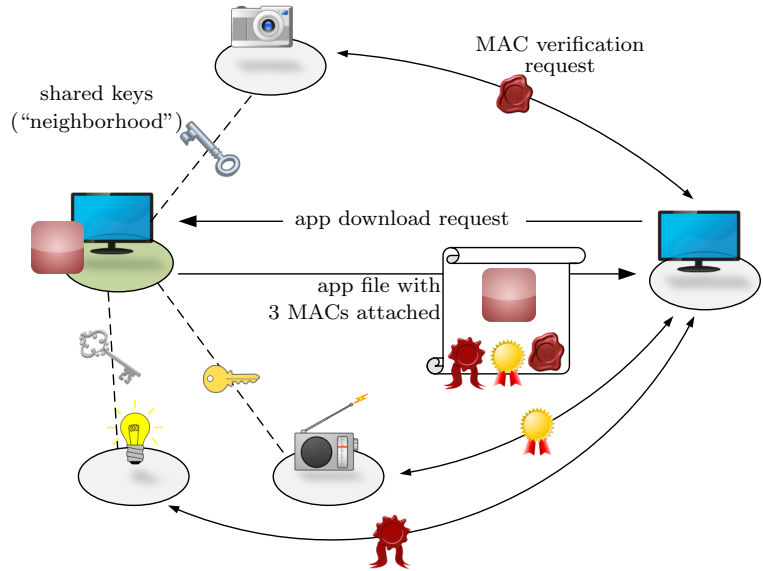

Suppose a user wants to install an app. The user has retrieved the app from
somewhere (not the app store), so that the source is distrusted (typically
indicated by a digital signature verification failure). What if the user –
for whatever reason – decides to need the app anyway? Where to get it from?
Apps retrieved from the internet directly are usually not subject to the
thorough screenings that app stores apply. In that case, the risk is fully
taken by the user. Why not rely on community knowledge and resources in that
case? The idea is the following: To retrieve the app (from elsewhere than the
app store but still from a trusted source), the user (device) $X$ performs
the following steps:

\begin{enumerate}
  \item\label{lbl:mpa-initialization} Call out to the community for ``Who's
      got `this app' ?''
  \item Some (perhaps many) devices may respond by sending a confirmation to
      have the app and sending the fingerprint (hash of the app file), upon
      which a set of $h_1,\ldots ,h_n$ arrives at the user.  Devices that
      are too old should be abandoned from that list e.g. if their key
      length (for calculating the MACs in step \ref{lbl:mpa-finalstep} is
      less than the current recommendations.
  \item The user $X$ goes for a majority vote and uses the hash that
      appears the most among $h_1,\ldots ,h_n$, following the hypothesis
      that the malware has not yet affected too large parts in the
      community. Let $h_i$ be that majority value received from the $i$-th
      respondent. Let the user having sent $h_i$ be called $B$ (as in
      Figure \ref{fig:mpa}).
  \item If any of the hashes mismatches another, then we have an indication
      of some malware potentially being around in the community.
  \item\label{lbl:mpa-finalstep} To retrieve the app, $X$ contacts $B$ to
      send the app with fingerprint $h_i$ as known from step 4. User $B$
      then runs multipath authentication to send an authentic version of
      the app. In Figure \ref{fig:mpa}, $B$ attaches MACs using the keys
      shared with its neighbors and sends the app file with the MACs to
      $X$. This one compares the fingerprint of the received app to match
      what it should be (namely $h_i$), and asks then (indicated) neighbors
      of $B$ to verify the hash of the app and the MAC attached to it. This
      prevents $B$ from sending a correct fingerprint but a malicious
      version of the app later (since the fingerprints would mismatch
      then), and assures $X$ the authenticity of $B$'s app packet.
\end{enumerate}

The plain protocol can be adapted towards a less stringent yet no less
informed behavior, in case that users are willing to accept a certain
residual risk in the retrieval process. If so, then quantifying that risk is
the major objective, and done as follows:

Let $N=\set{1,2,3,\ldots}$ represent the community, with physical member IDs
being numbers. To each member $i\in N$ of the community, we associate a trust
value based on the following intuition: if, in the above protocol, a user $X$
may query another user $B$ asking to respond to a MAC verification.
Ultimately, the user $X$ is interested in the trustworthiness of the
retrieved app; let us call this event $T\in\set{0,1}$, where $T=0$ is zero
trust and $T=1$ is full trust. Ultimately, we are interested in the
distribution of $T$ over the unit interval $[0,1]$, conditional on the
information available to $X$, i.e., our trust measure is $\tau=\Pr($app is
trustworthy$|$user $i$ says so$)$, or more compactly, $\tau=\Pr(T=1|R)$. Note
that $X$ polls only a subset of app sources/users $u_1,\ldots,u_n\in N$, so
that the probability space in which $T$ lives will not be partitioned by the
events $R_i$ associated with the peers in the protocol. To fix this, we
replace $T$ by $T_{X,B}$, as being the \emph{subjective} trust level that
user $X$ assigns to the app retrieved from user $B$, based on the information
available. This (conditional) random variable has its probability space
partitioned by the users that $X$ contacts, but the fix comes with the caveat
of the objective trust $T$ remaining out of reach, with only the subjective
trust $T_{X,B}$ assigned by $X$ being computable. The distribution of
$T_{X,B}$, however, follows from the law of total probability,
\begin{equation}\label{eqn:subjective-trust}
    \Pr(T_{X,B})=\sum_{i=1}^n \Pr(T_{X,B}=1|R_i)\Pr(R_i).
\end{equation}

Mechanism design is herein concerned with the question of \emph{why} a user
should be willing to honestly participate in the protocol. The
\emph{incentive} for user $i$ to respond correctly, by
\eqref{eqn:subjective-trust} is positively correlated to the subjective trust
that user $X$ obtains from asking user $i$. Since $i$ is, by construction,
$B$'s neighbor, whatever $i$ reports back depends on the trust that $i$ has
in $B$. But the situation is symmetric: $i$ has an incentive to answer
faithfully about $B$, since whichever information or app is ever retrieved
from $i$, $B$ is perhaps among the neighbors to be queried, so if $i$
refrains from responding or responds incorrectly, it may indirectly damages
its own reputation in the long run. The same symmetry, however, cannot
straightforwardly be used by a malicious $i$ to damage $B$'s reputation,
since there is still the set of other neighbors that may indicates $B$'s
honesty (recall that the protocol goes for the majority vote among all the
replies).

Practically, we have a triple of trust values, which $X$ maintains about the
community:
\begin{enumerate}
  \item The likelihood $\Pr(R_i)$ that user $i$ responds. Based on whether
      or not $i$ participated in the above protocol, $X$ can update the
      trust value for $i$ accordingly. If $i$ refuses to reply about $B$,
      $B$ can later refuse to tell about $i$, which is $i$'s indirect
      incentive to become active.
  \item The likelihood $\Pr(T_{X,B}=1|R_i)$ that user $i$'s reply was
      correct and helpful. This value is updated upon the outcome of the
      majority vote made along the protocol, as discussed before.
  \item The trust value $\Pr(T_{X,B}=1)$ computed from
      \eqref{eqn:subjective-trust}, based on the previous two items.
\end{enumerate}

This form of app retrieval is designed towards simplicity and security at the
same time: note that the community -- by collecting and comparing app
fingerprints -- continuously builds up some sort of threat intelligence. Note
that the hash in step \ref{lbl:mpa-initialization} and the MAC in step
\ref{lbl:mpa-finalstep} cryptographically link the app discovery and the app
download. Otherwise, a malicious device may present one (harmless) app at
step \ref{lbl:mpa-initialization} and another (malicious) one at step
\ref{lbl:mpa-finalstep} (cf. time of check to time of use -- TOCTTOU \cite{Mulliner.2012}).

The bandwidth increase is the primary price to pay for this scheme to work,
but remember that all we need to exchange are cryptographic hash values,
which are only a few bytes long. For a quick calculation: using a 224 Bit
SHA-3 checksum (state of the art) and with 10 peers contacted, the following
bandwidth is required: we have $10\times 224$ bit $\approx 2$ kBit for the
initial call out. Now, if the app sender upon request uses 10 of its
neighbors to verify its authenticity, we end up with a total of 2 kBit for
the 10 MACs, and another $10\times 2\times 224$ bit $\approx 4.5$ kBit for
all verifications. Thus, a total of less than 10 kBit of overhead is required
for 10 neighbors (more or less peers induce a proportional increase or
decrease respectively).

We finally stress that this entire scheme can be constructed on Layer 7,
i.e., no deep changes to the devices or their network stack is necessary. In
fact, a demonstration prototype of multipath authentication (in the natural
combination with multipath transmission) has been successful already.

The local management of keys is herein aligned with the app sandboxing
implemented on all commercial mobile phone platforms. General IoT devices are
expected to run the same (or at least similar) operating systems. Platforms
like Android enforce sandboxing towards preventing apps from accessing the
memory blocks assigned to other apps. Thus, the cryptographic keys are
essentially locally safe by logic access control by the operating system.

\section{Outlook}
The system discussed in this article already has close relatives up and
running to warn users about malicious websites or changed/outdated public key
certificates \cite{Wendlandt.2015,WOT.2015,vLoesch.2017}. Provided that such
technology does not itself exhibit unwanted behavior otherwise (like
unauthorized data collection), why not use similar technology in the IoT?

Hacking and threat intelligence are community actions, and so should security
and mutual protection be. Large (animal) populations protect themselves by
forming herds – and IoT devices can do similar things to harden the whole
community against external threats. We believe that a comprehensive security
concept should not exclusively rest on cryptographic mechanisms, but should
to a wide extent include incentive and credibility-driven mechanisms to let
users (devices) collaborate to the good of everyone. A combination of
mechanism design \cite{Borgers.2015}, game theory and cryptography can make a
start here, but the diversity of the IoT seems to call for a scientific
treatment with tools that are equally diverse.

\section*{References}

\bibliography{literature}

\end{document}